# Toward a Theory of Creative Inklings

Liane Gabora
University of British Columbia

It is perhaps not so baffling that we have the ability to develop, refine, and manifest a creative idea, once it has been conceived. But what sort of a system could spawn the initial seed of creativity from which an idea grows? This paper looks at how the mind is structured in such a way that we can experience a glimmer of insight or inkling of artistic inspiration.

## 1. MEMORY IS DISTRIBUTED BUT CONSTRAINED

Before we can explore how the contents of the mind are harnessed in a unique and innovative way, it is necessary to look briefly at how experiences are encoded in memory, and how they participate in subsequent experiences through contextual reminding events.

If the mind stored each experience in just one memory location as a computer does (Figure 1a), then in order for one experience to evoke a reminding of a previous experience, it would have to be *identical* to that previous experience. And since the space of possible experiences is so vast that no two ever *are* exactly identical, this kind of organization would be pretty useless. On the other hand, if any experience could activate any memory location (Figure 1b), the memory would be subject to crosstalk, a phenomenon wherein nonorthogonal patterns interfere. For the mind to be capable of evolving a stream of coherent yet potentially creative thought, the degree to which an experiences is *distributed* must lie between these extremes; that is, the size of the sphere of activated memory locations must fall within an intermediate range. A given instant of experience activates not just *one* location in memory, nor does it activate *every* location to an equal degree, but activation is distributed across many memory locations, with degree of activation decreasing with distance from the most activated one, which we call *k* (Figure 1c). The further a concept is from *k*, the less activation it not only *receives* from the current stimulus but in turn *contributes* to the next instant of experience, and the more likely its contribution is cancelled out by that of other simultaneously activated locations. A wide activation function means that locations relatively far from *k* still get activated; in other words, neurons have a lower activation threshold, so more fire in response to a given stimulus.

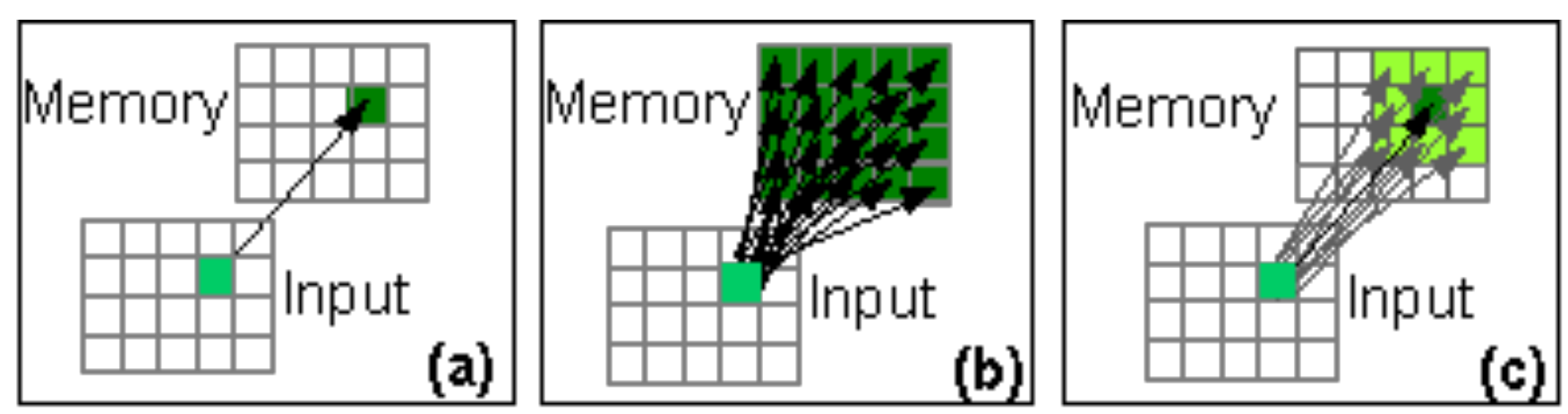

Figure 1. (a) A *one-to-one correspondence* between input and memory, as in a computer. (b) A *distributed* memory, as in some neural networks. (c) A *constrained distributed memory*, as in neural networks that use a radial basis function for the activation function. This is closest to how human memory works.

## 2. MEMORY IS CONTENT ADDRESSABLE

There is a systematic relationship between the *content* of an experience (not just as the subject matter, but the qualitative feel of it) and the memory locations where it gets stored (and from which material for the next instant of experience is evoked). In a computer, this kind of one-to-one correspondence is accomplished by assigning each possible input a unique address in memory. Retrieval is thus simply a matter of looking at the address in the address register and fetching the item at the specified location. The distributed nature of human memory prohibits this, but content addressability is still achievable as follows. The pattern of features (or phase relations) that constitutes a given experience induces a chain reaction wherein some neurons are inhibited and others excited. The address of a neuron is thus the pattern of excitatory and inhibitory synapses that make it fire. Since correlated qualia patterns get stored in overlapping locations, what emerges is that the system appears to retrieve experiences that are similar, or concepts that are relevant, to the current experience. As a result, the entire memory does not have to be searched in order for, for example, one painting to remind you of another.

## 3. A DISTRIBUTED, CONTENT ADDRESSABLE MEMORY CAN BE CREATIVE

A distributed, content addressable memory has advantages and disadvantages. Since stored items are 'smooshed' together in overlapping locations, at a high level of resolution, an item is never retrieved in *exactly* the form it was stored. Your new experience of it is reinterpreted in the context of similar experiences, and colored by events that have taken place since the last time you thought of it, as well as current stimuli, goals, and desires. Thus it is more accurate to think of the evoking process as *reconstruction* rather than retrieval**.**

Although this 'smooshing'/reconstruction is a source of inaccuracy, it enables the emergence of *abstractions*-concepts such as 'depth' with fewer dimensions than any of their instances. Abstractions unite stored experiences into an interconnected mental model of reality, or (from a first person standpoint) subjective worldview (Gabora 1998). The more abstract the concept, the greater the number of others potentially evoked by it. For instance, your concept of 'depth' is deeply woven throughout the matrix of concepts that constitute your worldview. It is latent in experiences as varied as 'deep swimming pool', 'deep-fried zucchini', and 'deeply moving book'; it derives its *existence* from these instances. The identification of an abstraction is a creative act. Often, however, when we think of creativity, we think of the invention of artifacts that merge lower-dimensional entities into something more *complex* than its constituents; for example, a dance is more than its steps. The 'smooshed' nature of human memory is the wellspring of *both* sorts of creativity. The more interwoven the mind is with abstractions, the more different ways of funneling an experience through the conceptual network, abstracting something new out of it, and manifesting the essence, or feel of it, through the constraints of an artistic medium (Gabora 2000).

## 4. THE CREATIVE INKLING

The current array of sensory stimuli can be viewed as a perturbation that impinges on the current spatio-temporal pattern of activated memory locations to create a *new* constellation of activated locations. You could say the perturbation *collapses* the conceptual network into a phenomenally manifested state; it reveals one 'slice' through the distribution of possibilities that was inherent in the conceptual network (Aerts & Gabora 1999). In a sense, the perturbation tests the integrity of a certain portion of the worldview, the size of the portion tested depending on the activation function. At the risk of mixing metaphors, you could say it's like throwing a ball against a wall and observing how it responds. The more flexible the material the ball is made of, the more it 'gives' when it makes contact. Similarly, the wider the activation function, the greater the portion of the worldview that makes contact with the world at that instant. Much as irregularities in the bounced ball cause its path to deflect, constrictions (repressed memories) or gaps (inconsistencies) in the 'collapsed' portion of the worldview may cause tension and thereby indicate a need for creative release, revision, or reconstruction.

(Of course, so long as the ball doesn't completely deflate and slide down the wall, you're doing fine. :-)

An *inkling*, then, is a collapse on an association or relationship amongst memories or concepts that, although their distribution regions overlap, were stored there at different times, and have never before been simultaneously perturbed, and evoked in the same collapse (Figure 2). Though it is a reconstructed blend, something never actually *experienced*, it can still be said to have been evoked from memory. It's like getting a 'bite' on many fishing rods at once, and when you reel them in you get a fish that is a mixture of the characteristics of the various fish that bit.

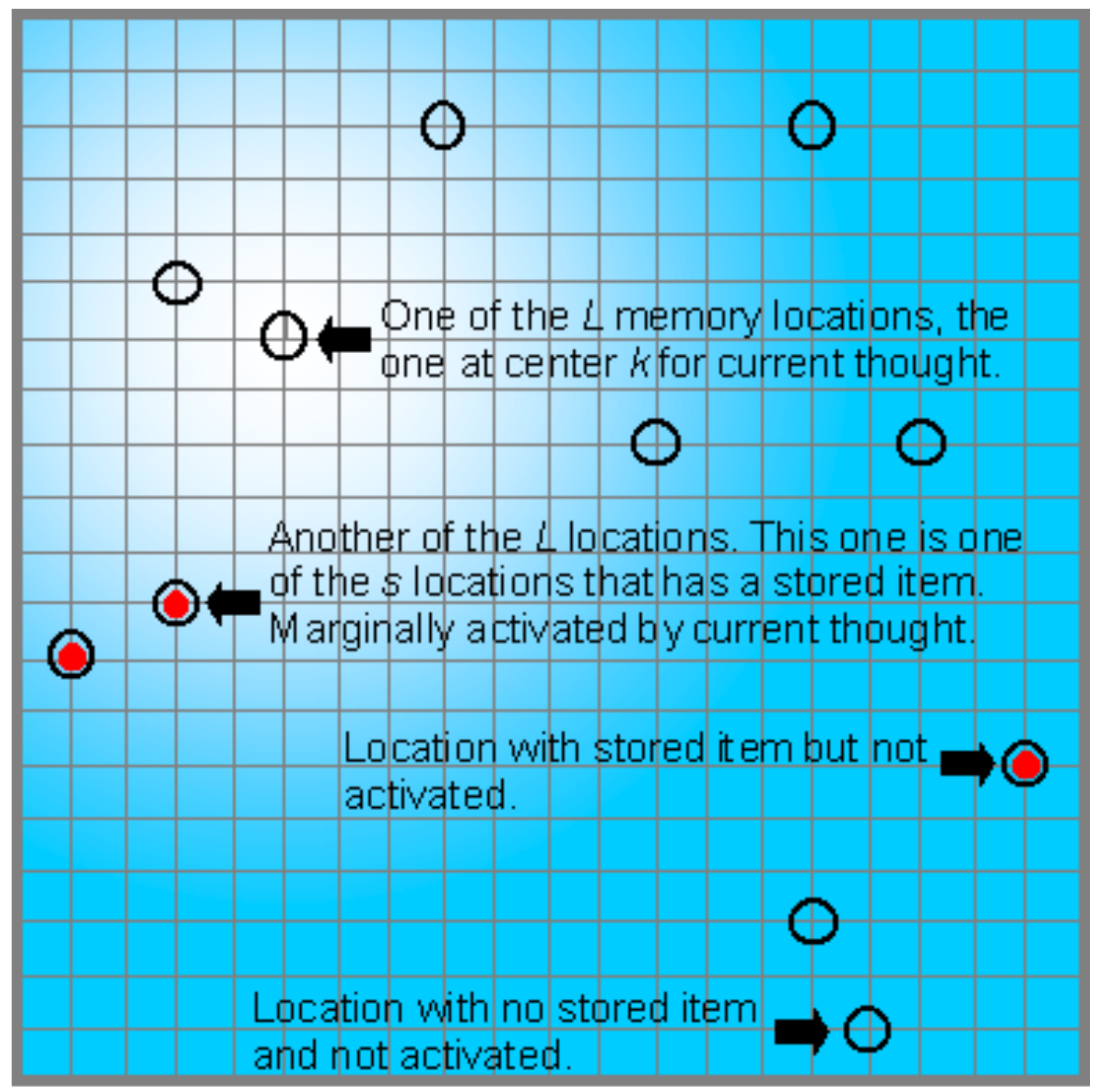

Figure 2. Schematic diagram of stimulus activating two dimensions of memory. Each vertex represents a *possible* memory location, and black dots represent *actual* locations. If many memories are stored in locations near *k*, they blend to generate the next experience.

**5. MENTAL STATES CONDUCIVE TO CREATIVITY**

If the activation function is large, a greater diversity of memory locations participate in the encoding of an instant of experience and release of 'ingredients' for the next instant. The more locations activated, the more *they* in turn activate, and so on; thus streams of thought tend to last longer. However, consecutive instants are less correlated because a concept from the periphery of the sphere of activated locations can pull the content of the next instant of experience far from its predecessor. *New* stimuli less readily command attention because they must compete with what has been set in motion by previous stimuli. If, however, something *does* manage to attract attention, it can get thoroughly assimilated into the matrix of abstractions, and thereby become increasingly decoupled from the stimulus that initially triggered it.

In the long run, since the relationship between one thought and the next could be so remote that a stream of thought lacks continuity, a large activation function would be untenable. However, in the short run, since there is a high probability of 'catching' new combinations of memories or concepts, it is conducive to creativity. The more features of the environment one attends or is sensitive to, the more memory locations potentially involved in its storage; thus, a large activation function may manifest as a state of defocused attention or heightened sensitivity. There is in fact experimental evidence that both defocused attention (Dewing & Battye 1971; Dykes & McGhie 1976; Mendelsohn 1976), and high sensitivity (Martindale & Armstrong 1974; Martindale 1977), including sensitivity to subliminal impressions (Smith & Van de Meer 1994) are associated with creativity.

One measure of creativity is the steepness of an individual's associative hierarchies (Martindale 1999; Mednick 1962). This is measured experimentally by comparing the number of words that individual generates in response to stimulus words on a word association test. Those who generate only a few words in response to the stimulus (e.g. 'chair' in response to 'table') have a *steep* associative hierarchy. Those who, after running out of the usual responses go on to generate unusual ones (e.g. 'elbow' in response to 'table') have a *flat* associative hierarchy. This is what one would expect with a wide activation function.

## 6. THE CREATIVE BEING AS CONDUIT

To make this more concrete let us take an example. Sometimes freestyle dancers find themselves 'doing moves', but unable to lose themselves to the music. The dancing feels and looks planned, does not touch the heart. Dance demands that the music be allowed to have its way with you, to forge new channels of expression through the constraints of your body, to use you to unearth something new. Perhaps this state, and its analog in other creative endeavors, derives from a state of defocused attention-thus widened activation function-such that *whatever* gets evoked, whether the logic of the association is apparent or not, can surface to the next instant of awareness.

## 7. PATTERN IN THE UNDERLYING REALITY MANIFESTS AS CONCEPTUAL LINKAGE DISEQUILIBRIUM

Let us look at the effect an inkling has on the conceptual network or worldview. The biological concepts of linkage equilibrium and disequilibrium are useful for gaining perspective on what is happening here. The closer together two genes are on a chromosome, the greater the degree to which they are *linked*. *Linkage equilibrium* is defined as random association amongst alleles of linked genes. Consider the following example:

*A* and *a* are equally common alleles of Gene **1**.
*B* and *b* are equally common alleles of Gene **2**.
Genes **1** and **2** are linked (nearby on same chromosome).

There are four possible combinations of genes **1** and **2**: *AB*, *Ab*, *aB*, and *ab*. If these occur with equal frequency, the system is in a state of linkage equilibrium. If not, it is in a state of linkage *dis*equilibrium. Disequilibrium starts out high, but tends to decrease over time because mutation and recombination break down arbitrary associations between pairs of linked alleles. However, at loci where this *does not* happen, one can infer that some combinations are fitter, or more adapted to the constraints of the environment, than others. Thus when disequilibrium does not go away, it reflects some structure, regularity, or pattern in the world.

What does this have to do with creativity? Like genes, the features of memories and concepts are connected through arbitrary associations as well as meaningful ones. We often have difficulty applying an idea or problem-solving technique to situations other than the one where it was originally encountered, and conversely, exposure to one problem-solving technique interferes with ability to solve a problem using another technique (Luchins 1942). This phenomenon, referred to as mental set, plays a role in cultural evolution analogous to that of linkage in biological evolution. To incorporate more subtlety into the way we carve up reality, we must first melt away arbitrary linkages amongst the discernable features of memories and concepts, thereby increasing the degree of equilibrium. As we destroy patterns of association that exist because of the historical contingencies of a particular domain, we pave the way for the forging of associations that reflect genuine structure in the world of human experience which may manifest in several or perhaps all domains. This needn't be an *intellectual* process. For example, one might have a sudden glimmer of insight into how the *feeling* of a particularly emotional experience could be extricated from the specifics of that experience, and re-manifest itself as, say, a piece of music.

**ACKNOWLEDGEMENTS**

I would like to acknowledge the support of the Center Leo Apostel and Flanders AWI-grant Caw96/54a. And I guess I have to thank my cat Inkling for inspiration.